\documentclass[11pt,epsf,letterpaper]{article}%
\usepackage{color}
\usepackage{amsmath}
\usepackage{amsfonts}
\usepackage{verbatim}
\usepackage{amssymb}
\usepackage{graphicx}
\usepackage{epstopdf}
\usepackage{mathrsfs}%
\providecommand{\U}[1]{\protect\rule{.1in}{.1in}}

\renewcommand{\b}[1]{\bar{#1}}
\renewcommand{\t}[1]{\tilde{#1}}
\begin{document}

\title{Hairy planar black holes in higher dimensions}
\author{$^{(1)}$Andr\'{e}s Ace\~{n}a, $^{(2,3)}$Andr\'{e}s Anabal\'{o}n, $^{(4)}$Dumitru
Astefanesei, and $^{(5,6)}$Robert Mann\\\\\textit{$^{(1)}$Instituto
de Ciencias B\'asicas, Universidad Nacional de Cuyo, Mendoza,
Argentina.} \\\\\textit{$^{(2)}$Departamento de Ciencias, Facultad
de Artes Liberales y}\\\textit{Facultad de Ingenier\'{\i}a y
Ciencias, Universidad Adolfo Ib\'{a}\~{n}ez, Vi\~{n}a del
Mar.}\\\\\textit{$^{(3)}$ Universit\'{e} de Lyon, Laboratoire de
Physique, UMR 5672, CNRS,}\\\textit{\'{E}cole Normale Sup\'{e}rieure
de Lyon}\\\textit{46 all\'{e} d'Italie, F-69364 Lyon Cedex 07,
France.} \\\\\textit{$^{(4)}$ Instituto de F\'\i sica, Pontificia
Universidad Cat\'olica de Valpara\'\i so,} \\\textit{ Casilla 4059,
Valpara\'{\i}so, Chile} \\\\\textit{$^{(5)}$ Department of Physics
and Astronomy, University of Waterloo,}\\\textit{Waterloo, Ontario,
Canada N2L 3G1.}\\\\\textit{$^{(6)}$ Perimeter Institute, 31
Caroline Street North Waterloo, Ontario Canada N2L 2Y5.}}

\maketitle

\begin{abstract}
We construct exact hairy planar black holes in D-dimensional AdS gravity.
These solutions are regular except at the singularity and have stress-energy
that satisfies the null energy condition. We present a detailed analysis of their
thermodynamical properties and show that the first law is satisfied.  We also
discuss these solutions in the context of AdS/CFT duality and construct  the
associated c-function.

\end{abstract}

\section{Introduction.}
The existence of scalar fields has been long expected from a theoretical perspective (e.g., they are amongst
the basic constituents of string theory). Since the discovery of the Higgs particle \cite{Aad:2012tfa}, it is
clear that scalar fields play an important role in physics. They may also play an important role for
understanding the black hole physics, in particular for higher dimensional black hole solutions of
supergravity, which is the low energy effective theory of string theory.

Motivated by AdS/CFT correspondence of Maldacena \cite{Maldacena:1997re},
 extensive has been invested in constructing asymptotically AdS solutions to Einstein gravity.
However, analytic hairy black hole solutions in Einstein-Dilaton theories are
rare. In \cite{Acena:2012mr}, we have constructed a family of \textit{exact}
hairy black hole solutions with planar horizons in five dimensional AdS
gravity. The extension of this analysis to higher dimensions is the goal of
this paper. We explicitly write down the solutions in a compact form and
discuss their thermodynamical properties.

Indeed, it is well known that hairy asymptotically AdS solutions are relevant in the
context of applied holography, e.g. for `bottom-up models' (see, e.g.,  \cite{Gubser:2008ny}).
However, many of these solutions
(for non-trivial scalar potentials) were generated numerically only and so the
physics of these solutions in the context of holography can be just partially
investigated. Exact solutions in gravity theories with scalar fields (and
developing techniques for finding these solutions) are very useful because one
can directly study some of their generic features. Since the AdS/CFT correspondence is
a weak/strong type of duality, by studying the classical gravity in the bulk
one can obtain important information about the strong regime of the dual field
theories, which admit gravity duals. We show that the null energy condition
(which is the relevant energy condition in AdS) is satisfied and so we expect
that the boundary theory is well defined.

The emphasis will be on classical properties of the solutions rather than
their implications for the dual field theory. However, even if we do not know
all the details of the dual field theory, we provide some computations on the
gravity side (for example  the c-function) that are important for the interpretation of our solution as an
RG flow. The dual gravity theory of a
D-dimensional RG flow between two fixed points is a geometry that interpolates
between two different AdS spaces (a domain wall) or a black hole at finite
temperature \cite{Akhmedov:1998vf}. In \cite{Freedman:1999gp} it was shown
that there is a deep connection between the c-theorem and null energy
condition in the context of domain wall solutions (see, \cite{Elvang:2007ba,
Astefanesei:2007vh} for a discussion in the context of black hole solutions).
Since the energy condition is satisfied for our solutions, the existence of a c-function that monotonically decreases, moving
in from the boundary, is therefore no  surprise.

Our paper is structured as follows. In Section \ref{sec2} we construct general $D$ dimensional black hole solutions to
the Einstein equations with scalar matter.  In Section \ref{sec3} we analyze some of the basic thermodynamics of our
solutions and in Section \ref{sec4} we consider specific examples in the context of AdS/CFT duality.  We close with some
concluding remarks and with two appendices that contain important details for obtaining the solutions and computing their mass.

\section{General solutions in $D$ dimensions}\label{sec2}

In this section we generalize the black hole solutions of \cite{Acena:2012mr}.
While exact $5$-dimensional planar black hole solutions were presented in
\cite{Acena:2012mr}, we are going to focus on other dimensions here and show
that the solutions can be nicely written in a compact form. By carefully
studying the asymptotic properties of these solutions, we will obtain the mass
of the scalar. In a particular case when a part of the potential vanishes, we
will also present a concrete superpotential associated to the scalar
potential. This case is particularly interesting to understanding of dynamics
in theories with domain walls \cite{Boonstra:1998mp}.

\subsection{General set-up}

We begin with the D-dimensional action:
\begin{equation}
I[g_{\mu\nu},\phi]=\frac{1}{2\kappa}\int_{M}d^{D}x\sqrt{-g}\left[  R-\frac
{1}{2}\partial_{\mu}\phi\partial^{\mu}\phi-V(\phi)\right]
\end{equation}
where $V(\phi)$ is the scalar potential and we use the convention $\kappa=8\pi
G_{D}$. Since we set $c=1=\hbar$, $\left[  \kappa\right]  =M_{P}^{-(D-2)}$
where $M_{P}$ is the D-dimensional Planck scale.

The equations of motion for the metric and dilaton are
\begin{equation}
E_{\mu\nu}=R_{\mu\nu}-\frac{1}{2}g_{\mu\nu}R-\frac{1}{2}T_{\mu\nu}^{\phi
}=0\,\,,\,\,\,\,\,\,\,\,\frac{1}{\sqrt{-g}}\partial_{\mu}\left(  \sqrt
{-g}g^{\mu\nu}\partial_{\nu}\phi\right)  -\frac{\partial V}{\partial\phi}=0
\end{equation}
where the stress tensors of the matter fields is
\begin{equation}
T_{\mu\nu}^{\phi}=\partial_{\mu}\phi\partial_{\nu}\phi-g_{\mu\nu}\left[
\frac{1}{2}\left(  \partial\phi\right)  ^{2}+V(\phi)\right]  . \label{enMom}%
\end{equation}

Next, as in \cite{Anabalon:2012ta, Acena:2012mr}, we use the following metric
ansatz:
\begin{equation}
ds^{2}=\Omega(x)\left[  -f(x)dt^{2}+\frac{\eta^{2}dx^{2}}{f(x)}+d\Sigma
^{2}\right]  \label{Ansatz}%
\end{equation}
where the parameter $\eta$ was introduced to obtain a dimensionless radial
coordinate $x$, $\Omega(x)$ is the conformal factor, and
\begin{equation}
d\Sigma^{2}=h_{ab}dx^{a} dx^{b},
\end{equation}
where we assume that greek indices run from $0$ to $D-1$ and latin indeces
from $2$ to $D-1$, with $x^{0}=t$, $x^{1}=x$. We also assume that the
$(D-2)$-dimensional metric $h_{ab}$ depends only on the coordinates $x^{a}$.

There are three independent (combinations of) equations of motion. One of them
in particular implies that $h_{ab}$ is an Einstein metric with constant scalar
curvature, that is
\begin{equation}
R[h]_{ab}=\frac{R[h]}{D-2}h_{ab},\qquad R[h]=constant.
\end{equation}
The relevant (combinations of) equations of motion are then
\begin{align}
E_{t}\,^{t}-E_{x}\,^{x}  &  =0\Longrightarrow\phi^{\prime2}=\frac{D-2}%
{2\Omega^{2}}\left[  3\left(  \Omega^{\prime}\right)  ^{2}-2\Omega
\Omega^{\prime\prime}\right] \label{eqphi}\\
E_{t}\,^{t}-\frac{1}{D-2}g^{ab}E_{ab}  &  =0\Longrightarrow f^{\prime\prime
}+\frac{D-2}{2\Omega}\Omega^{\prime}f^{\prime}+\frac{2\eta^{2}}{D-2}%
R[h]=0\label{eqf}\\
E_{t}\,^{t}+\frac{1}{D-2}g^{ab}E_{ab}  &  =0\Longrightarrow V=-\frac
{D-2}{2\eta^{2}\Omega^{2}}\left[  f\Omega^{\prime\prime}+\frac{D-4}{2\Omega
}f\left(  \Omega^{\prime}\right)  ^{2}+\Omega^{\prime}f^{\prime}\right]
+\frac{R[h]}{\Omega}. \label{eqV}%
\end{align}
where the derivatives are with respect to $x$. For a more detailed derivation
of the equations of motion and its integration see Appendix \ref{EOM}.

\subsection{Solutions}

In this section we are going to present the exact solutions obtained for a
general class of non-trivial moduli potentials, but before we briefly review
some relevant geometric aspects of AdS spacetimes within AdS/CFT duality (see,
e.g., \cite{Emparan:1999pm}).

By choosing different foliations of AdS spacetime, the boundary can be
approached in different ways so that the holographic field theory is living on
different geometries or topologies. In other words, the choice of `radial'
coordinate is important for defining the family of surfaces that approach the
boundary as a limit. In the context of AdS/CFT duality, the radial coordinate
plays the role of the \textit{energy scale} in the dual field theory. That is,
the radial flow in the bulk geometry can be interpreted as the RG flow of the
boundary theory --- when we say that the field theory is `living on the
boundary' it means that the processes near the boundary control the short
distance (or UV) physics.

Let us now present the $5$-dimensional AdS spacetime with three different
foliations,
\begin{equation}
ds^{2}=-\left(  K + \frac{r^{2}}{l^{2}} \right)  dt^{2} + \frac{dr^{2}%
}{K+\frac{r^{2}}{l^{2}}} + r^{2}d\Sigma^{2}_{K} \label{ads}%
\end{equation}
where $K=\{+1, 0, -1\}$ for the spherical ($d\Sigma^{2}_{1}= d\Omega^{2}$),
toroidal ($d\Sigma^{2}_{0}= \sum\limits_{a=1}^{3}dx_{a}^{2}$), and hyperbolic
($d\Sigma^{2}_{-1} = dH^{2}$) foliations respectively.\footnote{Here,
$d\Omega^{2}$ and $dH^{2}$ are the `unit' metrics on the $D$-dimensional
sphere and hyperboloid, respectively.} The AdS radius $l$
 is related to the cosmological constant via $\Lambda=-(D-1)(D-2)/2l^{2}$.

In these coordinates the boundary is at $r\rightarrow\infty$, for which the
induced metric is
\begin{equation}
d\gamma^{2}=\frac{r^{2}}{l^{2}}(-dt^{2} + l^{2}d\Sigma^{2}_{K})
\label{boundary}%
\end{equation}
This is a conformal boundary and the asymptotic boundary geometry is related
to the background geometry on which the dual field theory lives by a conformal
transformation (to get rid of the divergent conformal factor $r^{2}/l^{2}$).

The neutral static black holes with different horizon topologies exist in AdS
and the corresponding metrics are
\begin{equation}
ds^{2}=-f_{K}(r)dt^{2} + \frac{dr^{2}}{f_{K}(r)} + r^{2}d\Sigma^{2}_{K}
\label{bhs}%
\end{equation}
where $f_{K}(r)=K-\frac{m}{r^{D-3}}+\frac{r^{2}}{l^{2}}$.

Obviously when the mass parameter vanishes ($m=0$) we obtain   AdS
spacetime with the corresponding foliation (\ref{ads}). It is also worth
emphasizing that the different AdS patches are related by diffeomorphisms (in
the bulk) that correspond to (singular) conformal transformations in the
boundary. However, it is clear that   black holes with different horizon
topologies are different and cannot be related by a change of coordinates.

Within AdS/CFT duality, a black hole in AdS is described as a thermal state of
the dual conformal field theory.\footnote{Since in the $k=-1$ case the the
boundary geometry is conformal to Rindler space, the hyperbolic black hole is
a dual description of a thermal Rindler states of the CFT in flat space.}
Since the dual field theory can live on different topologies, the gravity
results `should know' about the Casimir energy (when there is a non-trivial
scale in the boundary, e.g. the $k=1$ case). In what follows, we focus on the
planar case ($k=0$) for which there is no Casimir energy associated to the
dual field theory.

As in \cite{Anabalon:2012ta, Acena:2012mr}, we choose the conformal factor
\begin{equation}
\Omega(x)=\frac{\nu^{2}x^{\nu-1}}{\eta^{2}\left(  x^{\nu}-1\right)  ^{2}}
\label{conffac}%
\end{equation}
so that we can solve \eqref{eqphi} to get
\begin{equation}
\phi=l_{\nu}^{-1}\ln(x)\,\,\,\,\,\,\,\,\,,\,\,\,\,\,\,l_{\nu}^{-1}=\sqrt
{\frac{(D-2)(\nu^{2}-1)}{2}} \label{phi}%
\end{equation}
Note that with our normalization the scalar field is dimensionless. The
parameter $\nu$ labels different hairy solutions.

To obtain the metric function $f(x)$ we have to integrate \eqref{eqf}. As a
simplifying assumption we consider from now on $R[h]=0$. Then
\begin{equation}
f(x)=C_{1}+C_{2}\sum_{k=0}^{D-2}\binom{D-2}{k}(-1)^{D-k}\frac{x^{\frac
{\nu(2k-D+2)+D}{2}}-1}{\nu(2k-D+2)+D}. \label{f}%
\end{equation}
We are going to rename the constants $C_{1}$ and $C_{2}$ in order to make
explicit the asymptotics of the solution. The first thing to notice is that
from \eqref{conffac} we have that the conformal boundary is located at $x=1$.
We have $f(x)=C_{1}$ and as we are interested in $AdS$ asymptotics, we fix
$C_{1}=1/l^{2}$, where the cosmological constant is $\Lambda
=-(D-1)(D-2)/(2l^{2})$. We denote $C_{2}$ by $\alpha$ and so the function
$f(x)$ reads
\begin{equation}
f(x)=\frac{1}{l^{2}}+\alpha\sum_{k=0}^{D-2}\binom{D-2}{k}(-1)^{D-k}%
\frac{x^{\frac{\nu(2k-D+2)+D}{2}}-1}{\nu(2k-D+2)+D}. \label{f2}%
\end{equation}

The singularities of the metric (and the scalar field) are at $x=0$ and
$x=\infty$, but they are enclosed by an event horizon. We shall concern
ourselves with the range $x\in\lbrack1,\infty)$, for which the
scalar field is positive.  In our coordinate system  the scalar field is regular
at the horizon and diverges  at the singularity $x=\infty$.  We shall see that
this implies $\alpha<0$ in order that our solutions have physically reasonable
thermodynamic behaviour.

It is easy to see that, for any value of $\Lambda$, $\nu$, and $x_{+}$, there
is an $\alpha$ such that $f(x_{+})=0$. Indeed, this simply follows from the
fact that $f(x)$ is linear in $\alpha$. Therefore, there are black holes in an
open set of the parameter space. Moreover, a simple inspection of the function
(\ref{f}) shows that it is regular for every positive $x\neq0$ and $x\neq\infty$.

As a point of clarification, we should note that the \textquotedblleft%
$-1$\textquotedblright\ in the numerator on the second term of \eqref{f2} is
part of the integration constant. We chose to write it in this form because it
allows us to separate the constant that determines the asymtotic $AdS$
character of the metric and because it leads to the correct limit when
$\nu$ is chosen to be $\nu=\frac{D}{D-2-2k}$ for $k$ between $0$ and $D-2$,
which is
\begin{equation}
\lim_{\nu\rightarrow\frac{D}{D-2-2k}}\frac{x^{\frac{\nu(2k-D+2)+D}{2}}-1}%
{\nu(2k-D+2)+D}=\frac{1}{2}\ln x. \label{lnlim}%
\end{equation}

As expected, one can also check directly that at the boundary, $x=1$, where
the scalar field vanishes, the metric becomes $AdS$ foliated with planar
slices (up to a change of coordinates). Also, it is worth pointing out that
for $\nu=1$ we obtain the Schwarzschild-AdS solution.

The dilaton potential is obtained from \eqref{eqV} using \eqref{conffac},
\eqref{phi} and \eqref{f2},
\begin{align}
\label{POT1}V(\phi)  &  =-\frac{D-2}{4\nu^{2}}\Bigg\{e^{-l_{\nu}\phi
}\Bigg[\frac{1}{l^{2}}+\alpha\sum_{k=0}^{D-2}\binom{D-2}{k}(-1)^{D-k}%
\frac{e^{\frac{1}{2}[\nu(2k-D+2)+D]l_{\nu}\phi}-1}{\nu(2k-D+2)+D}\Bigg]\\
&  \Big[2D(\nu^{2}-1)+e^{\nu l_{\nu}\phi}\big(D+(D-2)\nu\big)(\nu+1)-e^{-\nu
l_{\nu}\phi}\big(D-(D-2)\nu\big)(\nu-1)\Big]\nonumber\\
&  +\alpha e^{\frac{D-2}{2}l_{\nu}\phi}\Big(e^{\frac{\nu}{2}l_{\nu}\phi
}-e^{-\frac{\nu}{2}l_{\nu}\phi}\Big)^{D-2}\Big[2-e^{\nu l_{\nu}\phi}
(\nu+1)+e^{-\nu l_{\nu}\phi}(\nu-1)\Big]\Bigg\}\nonumber
\end{align}

At the boundary $x=1$, where the scalar field vanishes, the potential and its
first two derivatives are
\[
V(0)=-\frac{(D-1)(D-2)}{l^{2}}\qquad\left.  \frac{dV}{d\phi}\right\vert
_{\phi=0}=0\qquad\left.  \frac{d^{2}V}{d\phi^{2}}\right\vert _{\phi=0}%
=-\frac{(D-2)}{l^{2}}%
\]
Therefore, the theory has a standard AdS vacuum $V(\phi=0)=2\Lambda$ for
$\phi=0$. The second derivative of the potential is in fact the mass of the
scalar and for $D=5$, as expected from the previous work \cite{Acena:2012mr}, we obtain
$m^2=-3/l^{2}$. The rank of the coordinate $x$
can be taken to be either $x\in(0,1]$ or $x\in\lbrack1,\infty)$. The scalar
field is negative in the first case, but positive in the other range. Since
the dilaton potential has no obvious symmetry we therefore cover it completely.

Furthermore the correct limit is attained if $\nu=\frac{D}{D-2-2k}$ for $k$
between $0$ and $D-2$, where the potential contains a linear term in $\phi$,
corresponding to the $\ln x$ term in $f$.

\section{Thermodynamics}\label{sec3}

\label{thermo} In what follows, we compute the relevant
thermodynamical quantities and show that  the first law is indeed
satisfied. Since these are planar black holes, the boundary is flat
and so there is no Casimir energy associated with the dual field
theory. We can therefore calculate the mass using the method of
Ashtekar, Das, and Magnon \cite{Ashtekar:1999jx} even though this method
does not contain information about the Casimir energy.
This method resembles the calculation of the Bondi mass and also of
the ADM mass\footnote{Note that AMD refers to Ashtekar-Magnon-Das and that
ADM refers to Arnowitt-Deser-Minser.} used for asymptotically flat spacetimes, but
in the context of asymptotically AdS spacetimes. First, the AMD
mass resembles the Bondi mass in the sense that a certain integral
in a cut of the conformal boundary gives rise to quantities that
measure the flux of energy-momentum across the boundary. In
particular, if there is no flux of energy-momentum then the mass of
the spacetime is conserved and given by the leading order in the
asymptotic expansion of the Weyl tensor. As the conformal boundary
is time-like, it can be approached not only in null directions but
also in space-like directions, which is in particular how the
calculation is done. In this sense it resembles the ADM mass,
with the difference that now what would be the equivalent of
space-like infinity is no longer a point and therefore the mass need
not be conserved (though it is since our solutions have a timelike Killing vector).

We present here the outline of the calculations and leave the
details for Appendix \ref{MassDetails}. First we need to consider a
conformal metric to $g_{\mu\nu}$ that is regular at $x=1$. As a
straightforward choice we take
\begin{equation}
 \b{g}_{\mu\nu}=\b{\Omega}^2 g_{\mu\nu},
\end{equation}
with
\begin{equation}
 \b{\Omega}=\Omega^{-\frac{1}{2}}.
\end{equation}
The mass is then
\begin{equation}
 M=\left.\frac{l}{8\pi G_D(D-3)}\oint_{\Sigma}\b{\varepsilon}^t\,_t d\b{\Sigma}_t\right|_{x=1},
\end{equation}
where $\b{\varepsilon}^\mu\,_\nu$
\begin{equation}
 \b{\varepsilon}^\mu\,_\nu=l^2\b{\Omega}^{3-D}\b{n}^\rho\b{n}^\sigma\b{C}^\mu\,_{\rho\nu\sigma}
\end{equation}
is the electric part of the
conformal Weyl tensor $\b{C}^\mu\,_{\nu\lambda\rho}$
\begin{equation}
 \b{C}^\mu\,_{\nu\lambda\rho}= C^\mu\,_{\nu\lambda\rho}=g^{\mu\sigma}W_{\sigma\nu\lambda\rho},
\end{equation}
and $\b{n}^\mu$ is the normal vector to the boundary
\begin{equation}
 \b{n}_\mu=\partial_\mu\b{\Omega},
\end{equation}
satisfying $\left.\b{\nabla}_\mu\b{n}_\nu\right|_{x=1}=0$.
Performing the calculations and using the field equations, we get
\begin{equation}
 \b{\varepsilon}^t\,_t=\frac{l^2\alpha\nu^{D-2}(D-3)(D-2)}{32\eta^{D+2}(D-1)}\Omega^{-\frac{9}{2}}\Omega'^3 f.
\end{equation}

To calculate $d\b{\Sigma}_t$ we need the volume element on the
hypersurface $x=constant$, which is
\begin{equation}
 \mbox{Vol}=f^\frac{1}{2}\sqrt{|h_{ab}|}dt\wedge dx^2\wedge\ldots\wedge dx^{D-1},
\end{equation}
so that
\begin{equation}
 d\b{\Sigma}_t=\langle\partial_t,\mbox{Vol}\rangle=f^\frac{1}{2}\sqrt{|h_{ab}|}dx^2\wedge\ldots\wedge dx^{D-1}.
\end{equation}
The mass is then
\begin{equation}
 M=-\frac{1}{32\pi G_D}\frac{\alpha\nu^{D-2}}{\eta^{D-1}}\frac{D-2}{D-1}V_\Sigma
\end{equation}
where $V_\Sigma$ is the volume of the $(D-2)$-dimensional manifold
$\Sigma$.

We can also calculate the temperature and entropy of the black
brane, which are
\begin{equation}
 T=\frac{|f'(x_+)|}{4\pi\eta}=-\frac{\alpha\nu^{D-2}}{8\pi\eta^{D-1}}|\Omega(x_+)|^{-\frac{1}{2}(D-2)}=-\frac{\alpha}{8\pi\eta}\frac{|x_+^\nu-1|^{D-2}}{x_+^{\frac{1}{2}(\nu-1)(D-2)}},
\end{equation}
\begin{equation}
 S=\frac{A}{4G_D}=\frac{(\Omega(x_+))^\frac{D-2}{2}V_\Sigma}{4G_D}=\frac{\nu^{D-2}x_+^{\frac{1}{2}(\nu-1)(D-2)}}{4G_D\eta^{D-2}|x_+^\nu-1|^{D-2}}V_\Sigma,
\end{equation}
and we see  that $\alpha <0$ in order for $M$ and $T$ to both be positive.

It is straightforward to check that the first law is satisfied. However, there is
a subtlety related to the coordinate system we use. Consider the first law
for similar coordinates in a simpler example, namely the planar
Schwarzschild AdS black hole:
\begin{equation}
ds^{2}=-(\frac{r^{2}}{l^{2}}-\frac{\mu }{r^{2}})dt^{2}+(\frac{r^{2}}{l^{2}}-%
\frac{\mu }{r^{2}})^{-1}dr^{2}+r^{2}d\Sigma
\end{equation}%
where $\mu $ is the mass parameter. With the following change of
coordinates
\[
r=\mu ^{\frac{1}{4}}x
\]
the metric becomes
\[
ds^{2}=-\mu ^{\frac{1}{2}}(\frac{x^{2}}{l^{2}}-\frac{1}{x^{2}})dt^{2}+(\frac{%
x^{2}}{l^{2}}-\frac{1}{x^{2}})^{-1}dx^{2}\mu ^{-\frac{1}{2}}+\mu ^{\frac{1}{2%
}}x^{2}d\Sigma
\]

In these coordinates the location of the horizon $x_{+}$ is
independent of the mass. Actually $x_{+}=l^{\frac{1}{2}}$ in this
example. The thermodynamical quantities in terms of $x_{+}$ can be
written as:

\begin{equation}
S = \frac{\mu ^{\frac{3}{4}}x_{+}^{3}Vol(\Sigma )}{4}
\,\,\,\,\,\,\,\, T = \frac{\mu ^{\frac{1}{4}}}{2\pi
}(\frac{x_{+}}{l^{2}}+\frac{1}{x_{+}^{3}}) \,\,\,\,\,\,\,\, M =
\frac{3\mu }{16\pi }Vol(\Sigma )
\end{equation}

To check the first law, it is only necessary to make the variation
with respect to $\mu$. Then, it is easy to check the first law
$T\delta S=\delta M$. Similarly, it is also straightforward to show
that the first law is satisfied for our solutions (by considering
the variation with respect to $\eta$ only.)

The AMD mass, as computed, includes no contribution from the
scalar field.\footnote{In some cases there is a contribution from the scalar
fields \cite{Henneaux:2006hk}, but since the first law is not affected,  we
are going to present a detailed analysis using different methods existent in the literature,
e.g. Hamiltonian mass and counterterm method, in a future work \cite{noi1}.} From
the metric expansion we see that the AMD mass coincides with the
gravitational mass (see, also, \cite{Chen:2005zj})
\begin{equation}
-g_{tt}=\frac{r^{2}}{l^{2}}+\frac{\alpha\nu^{2}}{3\eta^{3}r}+O(r^{-2})
\end{equation}
upon inspection of the subleading term in the large-$r$ expansion.

\section{Examples and $AdS/CFT$ duality}\label{sec4}

AdS/CFT duality has stimulated the study of AdS gravity in various
dimensions. In particular, AdS black hole solutions in five and seven
dimensions  are relevant for $AdS_{5}/CFT_{4}$ and $AdS_{7}/CFT_{6}$
dualities.  Furthermore  domain wall solutions in AdS are important in   so-called `fake supergravity' \cite{Skenderis:1999mm}.

In this section we  present some concrete examples and comment on
the significance of our solutions in the context of AdS/CFT duality. First, we
are going to obtain analytic domain wall solutions in D-dimensions and
explicitly construct the superpotential. We then show that the null energy
condition is satisfied for our solutions and construct a general c-function in
any dimension. We also present concrete black hole examples in $5$ and $7$ dimensions.

\subsection{Domain walls}

The scalar potential (\ref{POT1}) has two parts, one which is controlled by
the cosmological constant $\Lambda\sim\frac{1}{l^{2}}$ and the other by
$\alpha$ that is an arbitrary parameter with the same dimension as $\Lambda$,
namely $[\alpha]=[L^{-2}]$. When $\alpha$ vanishes, the solution becomes a
domain wall\footnote{This can be easily seen from the metric function
(\ref{f2}), which becomes $1/l^{2}$ in this limit. Then, the conformal factor
$\Omega(x)$ becomes the only relevant metric function that controls the
geometry and profile of the scalar.} and the potential can be rewritten as
\[
V(\phi)=-\frac{D-2}{4l^{2}\nu^{2}}\left[  e^{\phi l_{\nu}(\nu-1)}%
(\nu+1)\left(  (\nu+1)D-2\nu\right)  -e^{-\phi l_{\nu}(\nu+1)}(\nu-1)\left(
2\nu-(\nu-1)D\right)  +2e^{-\phi l_{\nu}}(\nu^{2}-1)D\right]
\]

In this particular case, we are able to explicitly write down a superpotential
associated with the potential above.  We obtain
\begin{equation}
V(\phi)=(D-2)\left(  \frac{dP(\phi)}{d\phi}\right)  ^{2}-\frac{(D-1)}{2}%
P(\phi)^{2}%
\end{equation}
where
\begin{equation}
P(\phi)=\frac{\sqrt{D-2}}{\sqrt{2}l\nu}\left[  (\nu+1)e^{\frac{\phi l_{\nu
}(\nu-1)}{2}}+(\nu-1)e^{-\frac{\phi l_{\nu}(\nu+1)}{2}}\right]
\end{equation}
\begin{equation}
\frac{dP(\phi)}{d\phi}=\frac{\sqrt{D-2}l_{\nu}(\nu^{2}-1)}{2\nu\sqrt{2}%
l}(e^{\frac{\phi l_{\nu}(\nu-1)}{2}}-e^{-\frac{\phi l_{\nu}(\nu+1)}{2}})
\end{equation}
In this case, the equations of motion can be rewritten as first order
equations, which is the fake supergravity method \cite{Skenderis:1999mm}.
Since this method requires the general structure of supergravity only to
lowest order in fermion fields, it is clear that it is far less restrictive
than true supergravity. These concrete solutions that satisfy first order
equations may be non-BPS solutions of true supergravity, though at this point
it is not clear to us if the potential can be embedded in string theory. On
the other hand, even if embedding in string theory is not possible, it is
clear that the theory has some resemblance to supergravity and we expect that
these solutions are linearly stable, similar to  previously known non-BPS solutions.

\subsection{Black hole solutions and c-function}

As a warm up exercise, we begin with a brief review of the 5-dimensional
planar black holes presented in \cite{Acena:2012mr}. These solutions can be
obtained from our general expressions for the following values of the
integration constants:
\begin{equation}
C_{1}=\frac{1}{l^{2}} \,,\,\,\,\,\,\,\,\,\, C_{2}=\frac{\alpha}{36\nu^{3}}%
\end{equation}
For concreteness, let us explicitly write down the relevant metric function
for $D=5$:
\begin{equation}
f(x)=-\frac{\Lambda}{6}+\alpha\left[  \frac{4}{3\left(  \nu^{2}-25\right)
\left(  9\nu^{2}-25\right)  }+\frac{x^{\frac{5}{2}}}{12\nu^{3}}\left(
\frac{x^{\frac{3\nu}{2}}}{3(3\nu+5)}+\frac{x^{-\frac{3\nu}{2}}}{3(3\nu
-5)}-\frac{x^{\frac{\nu}{2}}}{(\nu+5)}-\frac{x^{-\frac{\nu}{2}}}{(\nu
-5)}\right)  \right]
\end{equation}
Since their thermodynamic and holographic properties were discussed in
detail in \cite{Acena:2012mr}, we do not repeat the analysis here. Instead, we
present the asymptotics of these solutions and compare with the results of
\cite{Henneaux:2006hk} so that the analysis of these black holes is complete.

We change the coordinates
\begin{equation}
\ln x=\frac{1}{\eta r}-\frac{1}{2\eta^{2}r^{2}}-\frac{\nu^{2}-9}{24\eta
^{3}r^{3}}+\frac{\nu^{2}-4}{12\eta^{4}r^{4}} \label{form}%
\end{equation}
so that our metric matches the asymptotic form of \cite{Henneaux:2006hk}. We
obtain then
\begin{equation}
h_{rr}=-\frac{(\nu^{2}-1)l^{2}}{4\eta^{2}r^{4}}+...\,, h_{tt} =-\frac
{\alpha+\frac{3(9\nu^{2}-25)(\nu^{2}-25)}{10l^{2}}}{288\eta^{4}r^{2}}+...\,,
h_{mn} = \frac{(9\nu^{2}-25)(\nu^{2}-25)}{960\eta^{4}r^{2}}\delta_{mn}+...
\end{equation}
where $h$ is the asymptotic departure from locally AdS
spacetime as given in (\ref{ads}). A second important point that was
not discussed in \cite{Acena:2012mr} is the limit $\nu=5$ or
$\nu=\frac{5}{3}$. At first sight, it seems that the metric diverges
for these special cases. However, based on our general analysis in
Section $2$, we are going to show that, in fact, the solution (and
the corresponding potential) is well defined for $\nu=5$ (similarly,
one can prove that the solution is also well defined for
$\nu=\frac{5}{3}$). An explicit analysis of $\nu=5$ case (which
corresponds to $k=1$) shows that by computing \eqref{f2} using
\eqref{lnlim} gives
\begin{equation}
f(x)=\frac{1}{l^{2}}+\alpha\left(  \frac{x^{-5}-1}{10}+\frac{3}{2}\ln
x-3\frac{x^{5}-1}{10}+\frac{x^{10}-1}{20}\right)  ,
\end{equation}
The same result can be also obtained if we perform directly the integration of
\eqref{eqf} with this choice of $D$ and $\nu$. In this case, the situation
drastically changes in the sense that   logarithmic terms appear in the metric. Note
that we must have $\alpha<0$ for $f$ to have a single root (and hence for the solution to have
an event horizon). The same is true for the 7D solution below, provided  $\nu\geq 1.28$.

The $7$-dimensional class of black hole solutions have the metric function
\begin{equation}
f(x)=\frac{1}{l^{2}}+\alpha\Bigg(\frac{x^{\frac{5\nu+7}{2}}-1}{5\nu+7}%
-5\frac{x^{\frac{3\nu+7}{2}}-1}{3\nu+7}+10\frac{x^{\frac{\nu+7}{2}}-1}{\nu
+7}-10\frac{x^{\frac{-\nu+7}{2}}-1}{-\nu+7}+5\frac{x^{\frac{-3\nu+7}{2}}%
-1}{-3\nu+7}-\frac{x^{\frac{-5\nu+7}{2}}-1}{-5\nu+7}\Bigg)
\end{equation}
The scalar potential is again a combination of exponentials involving the
scalar:
\begin{align}
V(\phi)  &  = -\frac{5}{4\nu^{2}}\Bigg\{e^{-l_{\nu}\phi}\Bigg[\frac{1}{l^{2}%
}+\alpha\Bigg(\frac{e^{\frac{1}{2}(5\nu+7)l_{\nu}\phi}-1}{5\nu+7}%
-5\frac{e^{\frac{1}{2}(3\nu+7)l_{\nu}\phi}-1}{3\nu+7}+10\frac{e^{\frac{1}%
{2}(\nu+7)l_{\nu}\phi}-1}{\nu+7}\\
&  -10\frac{e^{\frac{1}{2}(-\nu+7)l_{\nu}\phi}-1}{-\nu+7}+5\frac{e^{\frac
{1}{2}(-3\nu+7)l_{\nu}\phi}-1}{-3\nu+7}-\frac{e^{\frac{1}{2}(-5\nu+7)l_{\nu
}\phi}-1}{-5\nu+7}\Bigg)\Bigg]\\
&  \Big[14(\nu^{2}-1)+e^{\nu l_{\nu}\phi}\big(7+5\nu\big)(\nu+1)-e^{-\nu
l_{\nu}\phi}\big(7-5\nu\big)(\nu-1)\Big]\\
&  +\alpha e^{\frac{5}{2}l_{\nu}\phi}\Big(e^{\frac{\nu}{2}l_{\nu}\phi
}-e^{-\frac{\nu}{2}l_{\nu}\phi}\Big)^{5}\Big[2-e^{\nu l_{\nu}\phi}%
(\nu+1)+e^{-\nu l_{\nu}\phi}(\nu-1)\Big]\Bigg\}
\end{align}

A similar analysis as in the $5$-dimensional case can be done, but we do not
want to repeat such an analysis here. Instead, we are going to compute a
c-function on the gravity side that has the right properties.  As in the famous
$AdS_{5}/CFT_{4}$ correspondence, the $7$-dimensional black hole solution is
expected to describe the strong coupling behaviour of certain finite
temperature $6$-dimensional QFTs as well as the holographic RG flows between
different $6$-dimensional CFTs. The conformal invariance in the bulk is broken
due to the non-trivial profile of the scalar.

First we would like to verify if the null energy condition (in the bulk) is
satisfied for our solutions. It is by now well known that violations of the
null energy condition will lead to superluminal propagation and instabilities
in the bulk \cite{Gao:2000ga} with corresponding violations in the holographic
dual theory (see, e.g., \cite{Kleban:2001nh}). The analysis can be done in
general and we show the validity of the null energy condition and present the
c-function in any dimension.

When formulated in asymptotically flat spacetimes the no hair theorems are a
statement of the fact that for theories with a scalar field potential
satisfying the condition $V(\phi)^{\prime\prime}\geq0,$ the stationary and
axisymmetric solutions are Ricci flat. Since this very condition can be
violated in AdS spacetime, a relevant question is what kind of energy
condition is satisfied by a generic scalar field theory. This question can be
easily determined for metrics of the form (\ref{Ansatz}).

The null energy condition demands that for any null vector $n^{\alpha
}n_{\alpha}=0$, the energy-momentum tensor must satisfy the inequality
$T_{\alpha\beta}n^{\alpha}n^{\beta}\geq0$. We work with an orthonormal frame
$e_{\mu}^{A}$ for which the energy density and pressure are diag$(\rho,
\,p_{x}, \,p_{a})=e_{\mu}^{A}e_{\nu}^{B}T^{\mu\nu}$, namely
\begin{equation}
\rho=\frac{g^{xx}\phi^{\prime2}}{2}+V,\qquad p_{x}=\phi^{\prime2}-\rho
\text{,}\qquad p_{a}=-\rho.
\end{equation}
It easily follows that the null energy condition it is satisfied: $\forall p$,
\,\,$\rho+p\geq0$.

Let us know turn to the computation of the c-function --- here we follow
closely \cite{Freedman:1999gp, Astefanesei:2007vh}. The idea of an RG flow for
quantum field theory is based on the intuition that a coarse graining removes
the information about the small scales. In the context of holography, since
there is a gradual loss of non-scale invariant degrees of freedom, there
should exist a c-function that is decreasing monotonically from the UV regime
(large radii in the dual AdS space) to the IR regime (small radii in the
gravity dual). A function with this property can be built on the gravity side
as follows.

A $D-$dimensional metric of the form
\begin{equation}
ds^{2}=-a^{2}(x)dt^{2}+b^{-2}(x)dx^{2}+c^{2}(x)d\Sigma
\end{equation}
has an Einstein tensor, $G_{\mu}^{\nu},$ such that%

\begin{equation}
\phi^{\prime2} = \left(  G_{x}^{x}-G_{t}^{t}\right)  b^{-2}(x) = (D-2)\left[
-c^{\prime\prime}+c^{\prime}\left(  \ln\left(  \frac{a} {b}\right)  \right)
^{\prime}\right]  =(D-2)\frac{c^{\prime}}{c}\left[  \ln\left(  \frac
{a}{bc^{\prime}}\right)  \right]  ^{\prime}%
\end{equation}
It is then straightforward to see that
\begin{equation}
C(x)=C_{0}\left(  \frac{a}{bc^{\prime}}\right)^{(D-2)}
\label{C}%
\end{equation}
is a monotonically increasing function of r for any positive constant $C_{0}$.
The power was chosen so that the c-function matches the entropy of the black
hole at the horizon --- since $c^{\prime}(r)$ acts as the conformal radius of
the surface $\Sigma$, the c-function should indeed scale as $(c^{\prime
}(r))^{D-2}$. One can also check that this is the right power by taking the domain
wall limit and compare with the c-function of \cite{Freedman:1999gp} (there, the c-function
was fixed by comparing it with the two-point function of the stress tensor).
When evaluated for our solution the c-function (\ref{C}) becomes
\begin{equation}
\label{C1}C(x) = C_{0}\left(  \frac{2\eta \Omega^{3/2}}{\Omega^{\prime}}\right)^{(D-2)}  = \mathcal{C}_{0}\left[  \frac{x^{\frac{\nu+1}{2}}}{x^{\nu}%
(\nu+1)+(\nu-1)}\right]  ^{(D-2)}%
\end{equation}
where we have absorbed some constants into $\mathcal{C}_{0}$.

Let us conclude this section with a discussion of the c-function for the hairy
and regular planar black hole. First, let's work with the general expression
(\ref{C}) and observe that for the usual planar AdS black hole, $a=b$ and
$c^{\prime}=1$ and so the c-function is a constant. This is what we expect
because the flow is trivial in this case, the black hole is a thermal state in
the dual theory. On the other hand, for the hairy black hole the flow is
non-trivial. This can be seen from (\ref{C1}) because the c-function is
constant just for $\nu=1$, which corresponds to the no-hair case.

\section{Conclusions}\label{sec5}

 In this paper, we have constructed a family of higher-dimensional
asymptotically AdS  hairy black branes. These are the first explicit
analytical examples of {\it neutral  regular} hairy planar black holes in
dimensions higher than five.\footnote{Using the techniques
presented in \cite{Anabalon:2013qua}, one could generalize these
solutions by including gauge fields and obtain new extremal and
non-extremal  black hole solutions \cite{noi}.} Our solutions are regular
outside of the horizon and have stress-energy that satisfies the
null energy condition. They can be straightforwardly generalized to solutions with
spherical or hyperbolic compact sections.

Settings where there is no supersymmetry or conformal symmetry to constrain
the dynamics could play an important role to studying real world systems where
these symmetries are absent. Therefore, studying the generic properties of
exact hairy solutions can be useful even if the underlying dynamics of the
dual field theory is poorly understood.

The scalar potential has two parts, one that vanishes at the boundary where
the scalar field also vanishes and another part that, as expected, becomes the
cosmological constant at the boundary (so that our solutions are asymptotically AdS).
The part that vanishes at the boundary is controlled by a parameter $\alpha$,
and when $\alpha=0$ the solution is a naked singularity/domain wall. From this
point of view, our solutions are also important because they provide concrete
examples of naked singularities that can be dressed with a horizon due to the
new non-trivial ($\alpha$-)terms/corrections in the potential.

We have also constructed a c-function  (\ref{C1}) for this class of solutions. This
result relies only on the structure of the field equations.
For the usual AdS planar black hole this quantity is constant, whereas for
the class of solutions we construct it is a non-trivial monotonically increasing
function of radial distance (equivalently, it is monotonically decreasing as
the scalar flows from the asymptotic AdS boundary to the interior).  Given that the null energy condition is satisfied
 for these solutions, this is no surprise; however it does indicate that our
 solutions have an interpretation in the dual theory as an RG flow.

 When $\alpha=0$ we can explicitly write down the superpotential and study
 the RG flow. However, in this case, since the solutions are domain walls, the
 singularity is not hidden by a horizon and the scalar flow  is from $\phi=0$ at
 the boundary $x=1$ to $\phi \to \infty$ at the singularity $x=+\infty$. One resolution
 of this problem is to consider star solutions in AdS \cite{Hubeny:2006yu} which do not have a
 horizon and their entropy is much smaller than the black hole entropy
 (similar solutions with different scalar field configurations are
 the AdS boson stars \cite{Astefanesei:2003qy}).

For future work, it will be interesting to find the phase diagram of these solutions (in
fact there are both, first- and second-order phase transitions), to obtain similar solutions
but in spacetimes with positive cosmological constant, and to generalize these solutions
by including gauge fields.

\section{Acknowledgments}

We would like to thank David Choque for helpful discussions.
Research of A.A. is supported in part by the FONDECYT grant 11121187
and by the CNRS project \textquotedblleft Solutions exactes en
pr\'{e}sence de champ scalaire\textquotedblright. The work of DA is
supported in part by the Fondecyt Grant 1120446. DA would also like
to thank Albert Einstein Institute, Potsdam for warm hospitality
during the last stages of this project.

\appendix

\section{Details on the equations of motion}\label{EOM}

For the metric we have the ansatz \eqref{Ansatz} and therefore the
Christoffel symbols are given by
\begin{eqnarray}
 \Gamma^t\,_{\nu\lambda} & = & \delta^t_{(\nu}\delta^x_{\lambda)}\frac{1}{\eta}\Bigg(\frac{\Omega'}{\Omega}+\frac{f'}{f}\Bigg),\\
 \Gamma^x\,_{\nu\lambda} & = & \delta^t_\nu\delta^t_\lambda \frac{f^2}{2\eta^2} \Bigg(\frac{\Omega'}{\Omega}+\frac{f'}{f}\Bigg)+\delta^x_\nu\delta^x_\lambda \frac{1}{2\eta^2}\Bigg(\frac{\Omega'}{\Omega}-\frac{f'}{f}\Bigg)-\delta^a_\nu\delta^b_\lambda h_{AB} \frac{f}{2\eta^2} \frac{\Omega'}{\Omega},\\
 \Gamma^a\,_{\nu\lambda} & = & \delta^b_\nu\delta^c_\lambda\Gamma[h]^a\,_{bc}+\delta^x_{(\nu}\delta^a_{\lambda)}\frac{1}{\eta}\frac{\Omega'}{\Omega},
\end{eqnarray}
where the prime denotes the derivative with respect to $x$ and
$\Gamma[h]^a\,_{bc}$ are the Christoffel symbols of $h_{ab}$.

The Riemann tensor is given by
\begin{eqnarray}
 R^t\,_{\nu\lambda\rho} & = & \delta^x_\nu\delta^x_{[\lambda}\delta^t_{\rho]}\frac{1}{\eta^2}\left(\frac{\Omega''}{\Omega}+\frac{f''}{f}-\frac{\Omega'^2}{\Omega^2}+\frac{\Omega'f'}{\Omega f}\right)+h_{ab}\delta^a_\nu\delta^b_{[\lambda}\delta^t_{\rho]}\frac{f}{2\eta^2}\frac{\Omega'}{\Omega}\left(\frac{\Omega'}{\Omega}+\frac{f'}{f}\right),\\
 R^x\,_{\nu\lambda\rho} & = & -\delta^t_\nu\delta^t_{[\lambda}\delta^x_{\rho]}\frac{f^2}{\eta^3}\left(\frac{\Omega''}{\Omega}+\frac{f''}{f}-\frac{\Omega'^2}{\Omega^2}+\frac{\Omega'f'}{\Omega f}\right)+h_{ab}\delta^a_\nu\delta^b_{[\lambda}\delta^x_{\rho]}\frac{f}{\eta^3}\left(\frac{\Omega''}{\Omega}-\frac{\Omega'^2}{\Omega^2}+\frac{\Omega'f'}{2\Omega f}\right) \\
 R^a\,_{\nu\lambda\rho} & = & R[h]^a\,_{bcd}\delta^b_\nu\delta^c_{[\lambda}\delta^d_{\rho]}-\delta^t_\nu\delta^t_{[\lambda}\delta^a_{\rho]}\frac{f^2}{2\eta^2}\frac{\Omega'}{\Omega}\left(\frac{\Omega'}{\Omega}+\frac{f'}{f}\right)\\
 && +\delta^x_\nu\delta^x_{[\lambda}\delta^a_{\rho]}\frac{1}{\eta^2}\left(\frac{\Omega''}{\Omega}-\frac{\Omega'^2}{\Omega^2}+\frac{\Omega'f'}{2\Omega f}\right)+h_{bc}\delta^b_\nu\delta^c_{[\lambda}\delta^a_{\rho]}\frac{f}{2\eta^2}\frac{\Omega'^2}{\Omega^2},
\end{eqnarray}
where $R[h]^a\,_{bcd}$ is the Riemann tensor of $h_{ab}$. The Ricci
tensor and Ricci scalar are
\begin{eqnarray}
 R_{\nu\rho} & = & \delta^t_\nu\delta^t_\rho \frac{f^2}{2\eta^2}\left[\frac{\Omega''}{\Omega}+\frac{f''}{f}+\frac{D-4}{2}\frac{\Omega'^2}{\Omega^2}+\frac{D}{2}\frac{\Omega'f'}{\Omega f}\right] \\
 && -\delta^x_\nu\delta^x_\rho\frac{1}{2}\left[(D-1)\frac{\Omega''}{\Omega}+\frac{f''}{f}-(D-1)\frac{\Omega'^2}{\Omega^2}+\frac{D}{2}\frac{\Omega'f'}{\Omega f}\right]\\
 && -h_{ab}\delta^a_\nu\delta^b_\rho \frac{f}{2\eta^2} \left[\frac{\Omega''}{\Omega}+\frac{D-4}{2}\frac{\Omega'^2}{\Omega^2}+\frac{\Omega'f'}{\Omega f}\right]+R[h]_{ab}\delta^a_\nu\delta^b_\rho,\\
 R & = & \frac{R[h]}{\Omega}-\frac{1}{\eta^2}\frac{f}{\Omega}\left[(D-2)\frac{\Omega''}{\Omega}+\frac{f''}{f}+\frac{1}{4}(D-1)(D-6)\frac{\Omega'^2}{\Omega^2}+(D-1)\frac{\Omega'f'}{\Omega f}\right]
\end{eqnarray}
where $R[h]_{ab}=R[h]^c\,_{acb}$ and $R[h]=h^{ab}R[h]_{ab}$.

The energy-momentum tensor is given by \eqref{enMom}, and using the
metric expression we have
\begin{equation}
 T^\phi_{\mu\nu}=\delta^t_\mu\delta^t_\nu\frac{f^2}{2}\Bigg(\frac{\phi'^2}{\eta^2}+2V\frac{\Omega}{f}\Bigg)+\delta^x_\mu\delta^x_\nu\frac{\eta^2}{2}\Bigg(\frac{\phi'^2}{\eta^2}-2V\frac{\Omega}{f}\Bigg)-h_{ab}\delta^a_\mu\delta^b_\nu\frac{f}{2}\Bigg(\frac{\phi'^2}{\eta^2}+2V\frac{\Omega}{f}\Bigg).
\end{equation}

The Einstein equations are then
\begin{eqnarray}
\label{eqtt}  E_{tt} & = & \frac{f}{2}R[h]-\frac{f^2}{2\eta^2}(D-2)\Bigg(\frac{\Omega''}{\Omega}+\frac{1}{4}(D-7)\frac{\Omega'^2}{\Omega^2}+\frac{1}{2}\frac{\Omega'}{\Omega}\frac{f'}{f}\Bigg) - \frac{f^2}{4}\Bigg(\frac{\phi'^2}{\eta^2}+2V\frac{\Omega}{f}\Bigg)=0,\\
\label{eqxx} E_{xx} & = & -\frac{\eta^2}{2f}R[h]+\frac{D-2}{8}\frac{\Omega'}{\Omega}\Bigg((D-1)\frac{\Omega'}{\Omega}+2\frac{f'}{f}\Bigg)-\frac{\eta^2}{4}\Bigg(\frac{\phi'^2}{\eta^2}-2V\frac{\Omega}{F}\Bigg)=0,\\
  E_{ab} & = & R[h]_{ab}-\frac{1}{2}h_{ab}R[h] +h_{ab}\frac{f}{2\eta^2}\Bigg[(D-2)\frac{\Omega''}{\Omega}+\frac{f''}{f}+\frac{1}{4}(D-2)(D-7)\frac{\Omega'^2}{\Omega^2}+(D-2)\frac{\Omega'}{\Omega}\frac{f'}{f}\Bigg] \nonumber \\
  && +h_{ab}\frac{f}{4}\Bigg(\frac{\phi'^2}{\eta^2}+2V\frac{\Omega}{f}\Bigg)=0.
\end{eqnarray}
The last equation is equivalent to
\begin{equation}
 R[h]_{ab} = h _{ab}\frac{f}{(D-4)\eta^2}\Bigg[\frac{\eta^2}{2}\Bigg(\frac{\phi'^2}{\eta^2}+2V\frac{\Omega}{f}\Bigg)+(D-2)\frac{\Omega''}{\Omega}+\frac{f''}{f}+\frac{1}{4}(D-2)(D-7)\frac{\Omega'^2}{\Omega^2}+(D-2)\frac{\Omega'}{\Omega}\frac{f'}{f}\Bigg]
\end{equation}
and directly from this
\begin{equation}\label{eqab}
R[h] =
\frac{(D-2)f}{(D-4)\eta^2}\Bigg[\frac{\eta^2}{2}\Bigg(\frac{\phi'^2}{\eta^2}+2V\frac{\Omega}{f}\Bigg)+(D-2)\frac{\Omega''}{\Omega}+\frac{f''}{f}+\frac{1}{4}(D-2)(D-7)\frac{\Omega'^2}{\Omega^2}+(D-2)\frac{\Omega'}{\Omega}\frac{f'}{f}\Bigg].
\end{equation}
The l.h.s. of this equation depends only on $x^a$ and the r.h.s.
depends only on $x$, therefore $R[h]$ must be a constant. The same
can be deduced from the other two equations. If we just continue to
denote that constant by $R[h]$ we have that
\begin{equation}
 R[h]_{ab}=\frac{R[h]}{D-2}h_{ab}.
\end{equation}

So rearanging \eqref{eqtt}, \eqref{eqxx} and \eqref{eqab} gives
\begin{eqnarray}
\label{rear1} \frac{\phi'^2}{\eta^2}+2V\frac{\Omega}{f} & = & \frac{2}{f}R[h]-\frac{2(D-2)}{\eta^2}\Bigg(\frac{\Omega''}{\Omega}+\frac{1}{4}(D-7)\frac{\Omega'^2}{\Omega^2}+\frac{1}{2}\frac{\Omega'}{\Omega}\frac{f'}{f}\Bigg),\\
\label{rear2} \frac{\phi'^2}{\eta^2}-2V\frac{\Omega}{f} & = & -\frac{2}{f}R[h]+\frac{D-2}{2\eta^2}\frac{\Omega'}{\Omega}\Bigg((D-1)\frac{\Omega'}{\Omega}+2\frac{f'}{f}\Bigg),\\
\label{rear3} \frac{\phi'^2}{\eta^2}+2V\frac{\Omega}{f} & = &
\frac{2(D-4)}{D-2}\frac{R[h]}{f}-\frac{1}{\eta^2}\Bigg((D-2)\frac{\Omega''}{\Omega}+\frac{f''}{f}+\frac{1}{4}(D-2)(D-7)\frac{\Omega'^2}{\Omega^2}+(D-2)\frac{\Omega'}{\Omega}\frac{f'}{f}\Bigg).
\end{eqnarray}
If we add \eqref{rear1} and \eqref{rear2} and solve for $\phi'$ we
obtain the field equation for $\phi$, \eqref{eqphi}, namely
\begin{equation}\label{dphi}
 \phi'^2=\frac{D-2}{2\Omega^2}(3\Omega'^2-2\Omega\Omega'').
\end{equation}
If we substract \eqref{rear1} from \eqref{rear3} we obtain
\eqref{eqf} for $f$,
\begin{equation}\label{d2F}
 f''+\frac{D-2}{2\Omega}\Omega'f'+\frac{2\eta^2}{D-2}R[h]=0.
\end{equation}
Finally, adding \eqref{rear1} and \eqref{rear3} we have the equation
for $V$, \eqref{eqV},
\begin{equation}\label{potV}
 V=-\frac{D-2}{2\eta^2\Omega^2}\Bigg(f\Omega''+\frac{D-4}{2\Omega}f\Omega'^2+\Omega'f'\Bigg)+\frac{R[h]}{\Omega}.
\end{equation}

As stated we use now the conformal factor \eqref{conffac} and then
it is straightforward to solve \eqref{dphi},
\begin{equation}
 \phi=\sqrt{\frac{(D-2)(\nu^2-1)}{2}}\ln x,
\end{equation}
where the integration constant was chosen as to be $\phi=0$ at the
conformal boundary $x=1$. To obtain $f$ we notice that \eqref{d2F}
can be written as
\begin{equation}
 (\Omega^\frac{D-2}{2}f')'=-\frac{2\eta^2R[h]}{D-2}\Omega^\frac{D-2}{2}
\end{equation}
and therefore
\begin{equation}
 f=-\frac{2\eta^2R[h]}{D-2}\int^x\Omega(\t{x})^{-\frac{D-2}{2}}\Bigg(\int^{\t{x}}\Omega(\t{\t{x}})^\frac{D-2}{2}d\t{\t{x}}\Bigg)d\t{x}+c_2 \int^x\Omega(\t{x})^{-\frac{D-2}{2}}d\t{x}+c_1.
\end{equation}
At this point it is difficult to perform the double integral,
therefore we restrict ourselves to the case $R[h]=0$. Then we obtain
\eqref{f},
\begin{eqnarray}
\label{intf} f & = & c_1+c_2\frac{\eta^{D-2}}{\nu^{D-2}}\sum_{k=0}^{D-2}\binom{D-2}{k}(-1)^{D-k}\int^x\t{x}^{\nu k-\frac{1}{2}(\nu-1)(D-2)}d\t{x} \\
 & = & C_1+C_2\sum_{k=0}^{D-2}\binom{D-2}{k}(-1)^{D-k}\frac{x^{\frac{\nu(2k-D+2)+D}{2}}-1}{\nu(2k-D+2)+D},
\end{eqnarray}
where the integration constants are
\begin{eqnarray}
 C_1=c_1+c_2\frac{2\eta^{D-2}}{\nu^{D-2}}\sum_{k=0}^{D-2}\binom{D-2}{k}\frac{(-1)^{D-k}}{\nu(2k-D+2)+D},\hspace{1cm}C_2=c_2\frac{2\eta^{D-2}}{\nu^{D-2}},
\end{eqnarray}
and have been written in this form to make explicit the limit of the
function $f$ in case $\nu$ has been chosen as $\nu=\frac{D}{D-2-2k}$
for $k$ between $0$ and $D-2$, which is
\begin{equation}
 \lim_{\nu\rightarrow\frac{D}{D-2-2k}}\frac{x^{\frac{\nu(2k-D+2)+D}{2}}-1}{\nu(2k-D+2)+D}=\frac{1}{2}\ln x.
\end{equation}

We can now calculate $V(\phi(x))$ from \eqref{potV}, remembering
that in our case $R[h]=0$, and that from \eqref{intf}
\begin{equation}
 f'=c_2\Omega^{-\frac{D-2}{2}}=\frac{C_2(x^\nu-1)^{D-2}}{2x^{\frac{1}{2}(\nu-1)(D-2)}},
\end{equation}
we have
\begin{eqnarray}
V(\phi(x))  &  = & -\frac{D-2}{4\nu^{2}}\Bigg\{x^{-1}\Bigg[C_1+C_2\sum_{k=0}^{D-2}\binom{D-2}{k}(-1)^{D-k}\frac{x^{\frac{\nu(2k-D+2)+D}{2}}-1}{\nu(2k-D+2)+D}\Bigg]\\
&&  \Big[2D(\nu^{2}-1)+x^\nu\big(D+(D-2)\nu\big)(\nu+1)-x^{-\nu}\big(D-(D-2)\nu\big)(\nu-1)\Big]\\
&&  +C_2{x^{-\frac{1}{2}(\nu-1)(D-2)}}(x^\nu-1)^{D-2}\Big[2-x^{\nu}(\nu+1)+x^{-\nu}(\nu-1)\Big]\Bigg\}. %
\end{eqnarray}
Now we use \eqref{phi} to explicitly obtain $V(\phi)$ as a function
of $\phi$, which, after substituting $C_1$ and $C_2$ by $1/l^2$ and
$\alpha$ respectively, gives \eqref{POT1}.

\section{Details on the Weyl tensor and mass calculation} \label{MassDetails}

In this appendix we present the calculation of the mass of the
solution calculated following the ADM procedure
\cite{Ashtekar:1999jx}. For this we need the Weyl tensor, which is
defined as
\begin{equation}
 W_{\mu\nu\lambda\rho}=R_{\mu\nu\lambda\rho}-\frac{2}{D-2}(g_{\mu[\nu}R_{\rho]\nu}-g_{\nu[\lambda}R_{\rho]\mu})+\frac{2}{(D-2)(D-1)}Rg_{\mu[\lambda}g_{\rho]\nu}.
\end{equation}
For the metric \eqref{Ansatz} the expressions obtained in the
previous appendix lead to
\begin{eqnarray}
 W_{\mu\nu\lambda\rho} & = & \delta^t_{[\mu}\delta^x_{\nu]}\delta^t_{[\lambda}\delta^x_{\rho]}\frac{2}{D-1}\Omega\Bigg[(D-3)f''-\frac{2\eta^2}{D-2}R[h]\Bigg] \\
 && -\delta^t_{[\mu}\delta^a_{\nu]}\delta^t_{[\lambda}\delta^b_{\rho]}\frac{2}{D-2}\Omega f\Bigg[\frac{D-3}{D-1}\frac{f''}{\eta^2}h_{ab}+\frac{2}{D-1}R[h]h_{ab}-2R[h]_{ab}\Bigg] \\
 && +\delta^x_{[\mu}\delta^a_{\nu]}\delta^x_{[\lambda}\delta^b_{\rho]}\frac{2}{D-2}\frac{\Omega}{f}\Bigg[\frac{D-3}{D-1}f''h_{ab}+\frac{2\eta^2}{D-1}R[h]h_{ab}-2\eta^2R[h]_{ab}\Bigg]\\
 && +\delta^a_{[\mu}\delta^b_{\nu]}\delta^c_{[\lambda}\delta^d_{\rho]}\frac{2}{D-2}\Omega\Bigg[\frac{1}{D-1}\frac{f''}{\eta^2}h_{ad}h_{bc}+\frac{D-2}{2}W[h]_{abcd}\\
 && -\frac{4}{D-4}h_{ad}R[h]_{bc}+\frac{2(2D-5)}{(D-1)(D-3)(D-4)}R[h] h_{ad}h_{bc}\Bigg],
\end{eqnarray}
where $W[h]_{abcd}$ is the Weyl tensor of $h_{ab}$.


To calculate the mass we need a conformal metric which is regular at
the boundary $x=1$. As the divergence of $g_{\mu\nu}$ is only due to
the behaviour of the conformal factor we consider
\begin{equation}
 \b{g}_{\mu\nu}dx^\mu dx^\nu=\b{\Omega}^2 g_{\mu\nu}dx^\mu dx^\nu=-f(x)dt^2+\frac{\eta^2}{f(x)}dx^2+h_{ab}dx^adx^b,
\end{equation}
where
\begin{equation}
 \b{\Omega}=\Omega^{-\frac{1}{2}}.
\end{equation}
The ADM-mass is then
\begin{equation}
 M=\left.\frac{l}{8\pi G_D(D-3)}\oint_{\Sigma}\b{\varepsilon}^t\,_t d\b{\Sigma}_t\right|_{x=1},
\end{equation}
where $\b{\varepsilon}^t\,_t$ is the $tt$-component of the electric
part of the Weyl tensor,
\begin{equation}
 \b{\varepsilon}^\mu\,_\nu=l^2\b{\Omega}^{3-D}\b{n}^\rho\b{n}^\sigma\b{C}^\mu\,_{\rho\nu\sigma},
\end{equation}
the normal vector is obtained from
\begin{equation}
 \b{n}_\mu=\partial_\mu\b{\Omega}
\end{equation}
and the conformal Weyl tensor is
\begin{equation}
 \b{C}^\mu\,_{\nu\lambda\rho}= C^\mu\,_{\nu\lambda\rho}=g^{\mu\sigma}W_{\sigma\nu\lambda\rho}.
\end{equation}
One of the requirements to calculate the ADM-mass is that
$\left.\b{\nabla}_\mu\b{n}_\nu\right|_{x=1}=0$, which is
automatically satisfied by our choice of conformal factor. We have
then
\begin{equation}
 \b{\varepsilon}^\mu\,_\nu=\frac{l^2}{4\eta^4}\Omega^{\frac{1}{2}(D-9)}\Omega' f^2 g^{\mu\lambda}W_{\lambda x\nu x},
\end{equation}
and
\begin{equation}
 \b{\varepsilon}^t\,_t=-\frac{l^2}{4\eta^4}\Omega^{\frac{1}{2}(D-11)}\Omega'^2 f W_{txtx}=-\frac{l^2(D-3)}{8\eta^4(D-1)}\Omega^{\frac{1}{2}(D-9)}\Omega'^2 f f''.
\end{equation}
We use \eqref{d2F} and that with our choice of constants
\begin{equation}
 f'=\frac{\alpha\nu^{D-2}}{2\eta^{D-2}}\Omega^{-\frac{D-2}{2}},
\end{equation}
to finally get
\begin{equation}
 \b{\varepsilon}^t\,_t=\frac{l^2\alpha\nu^{D-2}(D-3)(D-2)}{32\eta^{D+2}(D-1)}\Omega^{-\frac{9}{2}}\Omega'^3 f.
\end{equation}

To calculate $d\b{\Sigma}_t$ we need the volume element on a
hypersurface $x=constant$, which is
\begin{equation}
 \mbox{Vol}=f^\frac{1}{2}\sqrt{|h_{ab}|}dt\wedge dx^2\wedge\ldots\wedge dx^{D-1},
\end{equation}
and then
\begin{equation}
 d\b{\Sigma}_t=\langle\partial_t,\mbox{Vol}\rangle=f^\frac{1}{2}\sqrt{|h_{ab}|}dx^2\wedge\ldots\wedge dx^{D-1}.
\end{equation}
Now we can perform the integration and obtain the mass
\begin{equation}
 M=\left.\frac{l^3\alpha\nu^{D-2}(D-2)}{256\pi\eta^{D+2}(D-1)}\Omega^{-\frac{9}{2}}\Omega'^3 f^\frac{3}{2} V_\Sigma\right|_{x=1}=-\frac{1}{32\pi G_D}\frac{\alpha\nu^{D-2}}{\eta^{D-1}}\frac{D-2}{D-1}V_\Sigma.
\end{equation}

\end{document}